\documentclass[12pt]{article}
\usepackage[dvips]{graphicx}
\usepackage[dvips]{epsfig}

\date{}

\begin{document}

\begin{center}

\textbf{Lagged Poincar\'{e} and auto-correlation analysis of Heart rate variability in diabetes
}

\vspace{1cm}

\textbf{S.K.Ghatak$^{a,*}$, B.Roy$^{b}$} \\
$^{a}$Department of Physics and Meteorology, $^{b}$School of Medical Science and Technology\\Indian Institute of Technology,Kharagpur-721392,India\\

\end{center}

\vspace{0.5cm}

\begin{abstract}

The heart rate variability (HRV) in diabetic human subjects, has been analyzed using lagged Poincar\'{e} plot, auto-correlation and the detrended fluctuation analysis methods. The parameters $SD1$, and $SD12 (= SD1/SD2)$ for Poincar\'{e} plot for diabetic are lower than that for non-diabetic subjects and reverse is case for $SD2$ for all lagged number (m). The slope and the curvature of the plot SD12 vs m is much reduced for diabetic subject. The scatter plot of two successive interbeat intervals points out smaller variability in diabetic heart. The detrended fluctuation exponent has a higher value for diabetic group. The auto-correlation function of the deviation of interbeat interval in diabetic group shows highly correlated pattern when compared to that of normal one. The study suggests that the curvature of $SD12$ and auto-correlation method appear to be better way to assess the alteration of regulatory system on heart dynamics in diabetic condition.

\end {abstract}

\vspace{1.0cm}


\vspace{0.5cm}

%\noindent\textbf{PACS}:71.10.Fd, 75.30.Mb

\vspace{2.0cm}

\noindent $^*$Corresponding author.

\noindent E-mail address: skghatak@phy.iitkgp.ernet.in

\newpage

\newcommand{\be}{\begin{equation}}
\newcommand{\ee}{\end{equation}}

\section{Introduction}
\label{Introduction}

The variability of human heart rate is well documented and unraveling of its complexity has drawn continuing attention. The cardiac oscillators are embedded in dynamic environment and are under the influence of regulatory biological system like autonomic nervous system (ANS). In order to cope with influencing signals some amount of plasticity in cardiac activity are therefore essential and this generates the variability of heart rate intervals. Heart Rate variability (HRV) is a measure of the fluctuations in the intervals between heartbeats, the RR interval. It reflects the cardiac autonomic regulation [1-3]. Diabetes, a common disease, may lead to the serious complication of autonomic diabetic neuropathy. Autonomic diabetic neuropathy is a major risk factor for cardiovascular mortality [4]. It leads to impaired regulation of blood pressure, heart rate and heart rate variability (HRV). It is associated with increased rate of cardiovascular disease such as heart attack, sudden cardiac death or Silent ischemia [5-9]. Early subclinical diagnosis of Autonomic Diabetic Neuropathy is difficult and may not be very sensitive. One of the common methods of detection is the Ewing Test Battery.  Ewing Test Battery consists of five tests and requires high cooperation from the patient. Its use is also limited by the fact that it cannot be used in patients with co-morbidities like existing heart disease or respiratory disease, which are contraindication for the Valsalva manoeuvre, an integrated part of the battery [10].Newer non invasive approaches like spectral analysis and Heart rate variability analysis are completely non invasive and are not associated with such limitations. They hold promise to provide a better diagnostic or prognostic tool for Diabetic Autonomic Neuropathy [11-14].

However, one has to be cautious while using the heart rate variability as conventionally used time and frequency domain parameters of HRV may not always represent the non-stationary characteristic of ECG rooted in nonlinear dynamics of the heart. A non linear method should be an appropriate agent to analyze the heart rate variability. Poincar\'{e} or Lorenz plot, a scatter plot of each RR interval as a function of the preceding one, is a non linear method with the ability to capture the non linear pattern of heart rate dynamics [15-18]. A recent approach with auto correlation technique also provides a new tool for this non linear system analysis [19]. In this study the time series of $RR_n$ -interval for number of diabetic and control subjects have been analyzed following the Poincar\'{e} plot and its modifications. Aim has been to reveal relative importance of the plot-characteristic parameters in differentiating a diabetic heart from non-diabetic one. We also stress that this investigation is of exploratory nature to determine the influence of the regulatory system on diabetic heart as elucidated by different kind of analysis of RR-interval. A marked difference between a diabetic and normal heart dynamics has been observed in $SD12$, correlation between successive difference in $RR_n$ -interval and the auto-correlation of fluctuation of $RR_n$ .

\section{Method}

We selected $23$ patients to carry out our research. We applied a standard exclusion criterion to ensure that any changes in HRV detected were due to the effect of the diabetes. A group of $23$ subjects was assigned as the control after matching the age distribution and other probable confounding parameters. We have maintained the principles outlined in the Declaration of Helsinki and a proper informed consent was given by every subject. All subjects were instructed to avoid caffeine, alcohol and heavy exercise the day before the study. The subjects were asked to relax for a period of $15$ minutes before we took the recording for $10$ minutes in supine position. One limitation of the study population was the non uniformity of medication among the study population.

\section{\textmd{Electrocardiogram (ECG) recording and HRV measurement}}

Three limb electrodes were placed on the appropriate limb to obtain the ECG signal. The sampling rate was $1000$Hz. Using A/D converter the ECG signal was recorded into a computer for analysis. Ectopic beats (if any) were selected visually and deleted manually from recorded data for $10$ minute interval. Then, RR intervals (in fact, NN intervals, i.e., normal-to-normal intervals) were measured and acquired through software written in Origin and Matlab.

\section{\textmd{Poincar\'{e} plot indices:}}

Poincar\'{e} plot, as described earlier, is basically a scatter plot of $RR_n$ and $RR_{n+1}$. When the plot is adjusted with an ellipse, the analysis provides three indexes: $SD1$, $SD2$ and ratio $SD12 =SD1/SD2$ [16]. $SD1$ is the standard deviation of instantaneous beat to beat interval variability and it is width of the ellipse and $SD2$ is the length of the ellipse. It has been shown that $SD1$ correlates with the short term variability of the heart rate and $SD2$ is a measurement of long term variability [17]. A strong correlation has been established between the high frequency powers of
Heart rate signals (modulated by parasympathetic nervous
system) to $SD2$. A similar correlation has been found between $SD2$ and both low and high frequency power (modulated by
both the parasympathetic and sympathetic nervous system)[20].
In addition to this conventional plot  ($RR_{n+1}$ vs $RR_n$) we also used the generalized Poincar\'{e}  plot  with different intervals i.e the m lagged Poincar\'{e} plot ( plot of $RR_{n+m}$ against $RR_n$). The concept of this m lagged plot emerged from the recognized notion that any given RR interval can influence up to eight subsequent RR intervals [21, 22].

\section{\textmd{Autocorrelation technique}}

Application of autocorrelation technique is a very recent idea and has enough potential to provide a new insight towards autonomic regulation of heart [19]. In this approach HRV is considered as the interaction between coupled oscillators of various frequencies [23]. It also reflects on the imbedded time scale of HRV. It has been taken as each of these time scale in the coupled oscillator is represented by a separate self-oscillator interacting with other oscillators with a particular physiological function. It enables us to count for the contribution of various systems like respiration, neurogenic, myogenic and endothelial towards the HRV. In this study we have applied the correlation technique to RR interval through the Origin software.

\section{Results and Discussions}

The plot of $RR_{n+m}$ with $RR_n$  for $m = 1,5$ and $9$ is depicted in Fig.1 for two  subjects of similar age but one is diabetic (left figure) and other is non diabetic and disease free (right figure). With an increase in the lag number the plot becomes more scattered with consequent increment in both width and length of the plot.For diabetic subject the width of the plot $RR_{n+m}$ vs $RR_n$ is smaller compared to that for non-diabetic person. SD1 and SD2 were calculated for lag m from the relation :SD1 = $(\Phi(m) - \Phi(0))^{1/2}$ and SD2 = $(\Phi(m) + \Phi(0))^{1/2}$ where  the  auto-covariance  function  $\Phi$(m) is  given by \\
\begin{center}
 $\Phi(M)$ =$E[(RR_N - \bar{RR}) (RR_{N+M} - \bar{RR})]$.
\end{center}
where $\bar{RR}$ is the mean $RR_n$[17].The extent of scatter in lagged Poincar\'{e} plot is quantified with the measure $SD1$, $SD2$ (both in sec.) and $SD12$, and the variation of these measure with lag number $m$ is presented in Fig.2 (points). Each point in the plot

 is obtained from calculation of the above parameters for individual subjects and then by taking the mean of that group. The values of $SD1$ and $SD12$ are higher in non-diabetic group compared to diabetic one, and the situation is reversed in case of $SD2$. The difference between the values of $SD12$ (points) in normal and diabetic subjects are more significant for different lag number. The differences between the values of all these parameters for two groups are statistically significant with a p value less than 0.001. In order to find the analytical relationship between $SD$'s and $m$ we used the method of Pade Approximant [24]. We assumed a simple form of the Pade approximant for $SD$'s as the ratio of polynomial in $m$ of degree one.

\begin{equation}
Y=\frac{a+bm}{c+dm}= \chi \frac{1+\beta m}{1+\gamma m}
\end{equation}

Here $Y = SD1, SD2$ or $SD12$ and $\chi = a/c$, $\beta = b/a$ and $\gamma = d/c$ were taken as new unknown parameters. An excellent fitting of the data with the equation (1) (solid line on the curve) was found with different set of $\chi$, $\beta$, $\gamma$ which were listed in table-1, and in all cases $R^2$ value was 0.999. In order to find relevance of these parameters in assessing health of cardio-vascular system we considered the eq. (1) for small m. In this limit the equation (1) can be approximated as $Y = C + L m + Q m^2$ where the slope L = $\chi (\beta - \gamma)$ and the curvature Q = - $\gamma L$.  So the slope and curvature of plot of $SD$ vs $m$ were determined by the fitted parameters $\chi$, $\beta$, $\gamma$.  Earlier, it was noted that the curvature of the plot ($SD12$-$m$) was significantly decreased for subject with cardio-vascular disease [21]. The values for L and Q as obtained from the fit of data by eq. (1) were given in table-1. The general features are that the slope is positive but curvature is negative for all parameters and curvature is nearly one order of magnitude smaller than the slope.

\begin{table}

\caption{ The value of parameters $\chi$, $\beta$, $\gamma$ obtained from fit of eq.(1) with respective value of $R^2$.The parameters L and Q are the coefficient of linear and quadratic terms in expansion of Y in terms of m. Values of $\chi$, L and Q for SD1 and SD2 are in second. }\

\begin{tabular}{|c|c|c|c|c|c|c|c|}

\hline
  % after \\: \hline or \cline{col1-col2} \cline{col3-col4} ...
   & a & $\chi$x$10^{-2}$& $\beta$x$10^{-2} $& $\gamma$x$10^{-2} $ & $R^2$ x$10^{-2}$& L x$10^{-3}$& Q x$10^{-4}$\\\hline
  SD1 & ND & $1.3\pm0.03$ & $39.1\pm2.0$ & $3.2\pm0.2$ & $99.9$ & $4.7\pm0.4$  & $1.5\pm0.2$  \\\hline
   & D & $1.00\pm0.02$& $38.2\pm1.4$ & $2.00\pm0.1$ & $99.9$& $3.6\pm0.08$  & $0.7\pm0.02$ \\\hline
 SD2 & ND & $3.2\pm0.06$  & $20.3\pm1.1$ & $3.5\pm0.2$ & $99.9$ & $5.4\pm0.04$ & $1.9\pm0.2$  \\\hline
   & D & $3.1\pm0.07$ & $26.4\pm1.6$ & $4.4\pm0.3$ & $99.9$ & $6.8\pm0.6$  & $3.0\pm0.5$  \\\hline
  SD12 & ND & $40.2\pm0.5$ & $25.0\pm1.8$ & $12.2\pm0.9$ & $99.9$ & $51.3\pm6.4$ & $62.7\pm2.4$ \\\hline
   & D & $33\pm0.3$ & $15.3\pm.8$ & $6.5\pm0.4$ & $99.9$ & $29\pm2.6$ & $18.9\pm2.8$ \\\hline

\end{tabular}

\end{table}

The cardiac oscillator is embedded in the dynamic environment regulated by ANS. This in turn leads to the correlation between interbeat variability.  The coherent nature of $RR_n$ interval can be assessed from the map of interval variation: $rr_{n+1}= \frac{RR_{n+2}-RR_{n+1}}{<RR_n>}$ vs $rr_n= \frac{RR_{n+1}-RR_n}{<RR_n>}$   where <$RR_n$>  represents the mean of $RR_n$. This plot reveals the correlation between the variabilities of three consecutive $RR$ intervals. Without any modulating effect the plot reduces to a point at origin whereas in presence of random input uniform distribution of point is expected. In Fig .3 the map of $rr_{n+1}$ -$rr_n$ for two subjects from each group (ND -control and D - diabetic) has been shown. The subjects of two groups are age matched.

Distinct differences are evident in the map. For diabetic patients the points are more crowded around origin. For diabetic person (Fig.3 D1) two successive decelerations (first quadrant) and acceleration (third quadrant) are as probable as combination of one acceleration and one deceleration (second and fourth quadrants). For D2 small asymmetry with more points in second and fourth quadrants is observed. In contrast, there exits more scatter about origin and higher asymmetry for normal person. Considering similar analysis for more subjects, the general features that emerge are high density of points around origin and smaller asymmetry for diabetic and more scattered with larger values of $rr_n$ and higher asymmetry for control group. It suggests that the influence of decelerating and accelerating mechanisms that act in different time frame on cardio-dynamics is substantially reduced in diabetic patients.

The reduced value of short term beat fluctuation $SD1$ indicates that the parasympathetic regulation in diabetic heart is weakened as expected. The higher value of $SD2$ in diabetic group means increased long term variability. Combination of these leads to higher ratio of $SD12$ in control group. The results for the Poincar\'{e} plot are more pointed when one examines the values of the slope (L) and curvature (- Q) of the plot. For diabetic group, L and -Q for $SD1$ and $SD12$ are smaller but their values for $SD2$ are higher compared to corresponding results for control group. The change in Q between two groups is more significant than the corresponding change in the values of L. In particular, the value of the curvature (Q) of $SD12$ in control group is found to be more than three times of Q for diabetic group. It may be mentioned that very small value of curvature is found in patient with cardio-vascular disease (21). Normally, $SD2$ is well correlated with both high (HF) and low (LF) frequency components of $RR_n$ fluctuation spectrum and $SD1$ is associated with HF.This data strongly suggest the reduction of parasympathetic activity and the over influence of sympathetic activity in the diabetic heart. Interestingly it also indirectly proves that sympathetic over influence has a strong correlation with cardiac morbidity [26,27].An increased $SD12$ is considered as good indicator of heart dynamics. The study supports notion of reduced sympatho-vagal balance in diabetes.

The strength of the correlation of heart rhythm can also be estimated by considering the auto-correlation of the deviation of $RR_n$ from its mean $<RR_n>$. The auto-correlation function $C(m)$  of a particular subject was calculated from $C(m)=\frac{\sum _{n=1}^N \Delta RR_{n+m}\Delta RR_n}{\sum _{n=1}^N \Delta {RR_n}^2}$ . Where the deviation $\Delta RR_n=RR_n-<RR_n>$  and N is total number of $RR_n$ interval. For representative result same two persons from two groups were considered and the results are displayed in Fig.4. The results for two diabetic subjects (D1 and D2 of Fig 3) and for two normal subjects (ND1 and ND2 of Fig.3) have been presented in upper and lower figures respectively. There also exits a distinctive feature in auto-correlation function for these two groups. For diabetic

subjects, the correlation function $C(m)$ decreases slowly (black and green curve of upper figure) with lag time. The time dependence was close to the sum of the two exponentials with superimposed small amplitude oscillation of low frequency. The presence and importance of the correlated pattern of beat interval was assessed by randomizing the RR-interval of each subject. The actual time series of RR- interval was shuffled using Matlab programme and the auto-correlation function of shuffled data (red and blue) were plotted in same Fig.4. The function has very small correlation time that characterizes the exponential decay of $C(m)$ . This indicates that a long time memory effect exists in diabetic condition. On the other hand the function $C(m)$ for actual time series of normal persons are found to have exponential fall with smaller correlation time compared to that for diabetic cases. The correlation function for normal subject could be fitted with form $C(m) \approx exp(-m/\Gamma)[a+b\cos(\omega m)]$. However, the auto-correlation functions of shuffled data of all subjects (2-diabetics and 2-non-diabetics) were nearly identical. The memory time ( $ \Gamma x<RR> $) was much smaller compared to the diabetic subjects.  The properties of $\Delta RR_n$  can also be characterized by the probability distribution function P ($\Delta RR_n$). In (Fig.5) the result of P ($\Delta RR_n$) is given for D1 and

ND1 subject. For diabetic person the probability distribution is almost symmetrical and fits with a Gaussian fit ($R^2 =0.93$) (full curve) with width $= 0.023$. For normal person the probability distribution P is asymmetrical with positive mean and higher width $= 0.036$ as obtained by the Gaussian fit ($R^2 =0.93 $).

Another way to assess long term correlation in RR-time sequence is based on the Detrended Fluctuation Analysis (DFA) [25]. The measure of correlation was given by a scaling exponent ($\alpha$) of the fluctuation function $F(\tau)\approx \tau^\alpha$ . The computation of fluctuation function  $F(\tau)$  was done in the following way. For a given time sequence $R(t_i)$, $t_i = i \delta t$ where $\delta t$ is characteristic time interval for the sequence and  i=1,N , an integrated time series $r(t_i)$ was defined as $r(t_i)=\sum_j^i [R(t_j)-<RR>]$, $i=1, N$ . The integrated series was divided into boxes of equal size of time  $\tau=n\delta t$ and linear function was used to fit box data. The fluctuation function $F(\tau)$  was calculated as root mean square fluctuations relative to the linear trend and $ \alpha $ was obtained from fitting power law behaviour. It has been observed that acceptable estimate of the scaling exponent $ \alpha $ (from DFA) can be obtained from analysis of data sets of length 256 samples or greater (equivalent to approximately 3.5 min for RR data at a heart rate of 70 bpm). The analysis of RR data for period of 10 min time interval was therefore expected to provide an adequate measure of the scaling exponent. The results are depicted in Fig.6 for two groups. The mean value of $ \alpha= 0.88 pm .17$    for normal person is smaller than that  $\alpha= 1.02 \pm .13$ for diabetic with $ p<0.001 $.

\section{Conclusions}

Diabetic condition is marked by much lower value of $SD12$ for all lag number. The data on parameters $SD1$, $SD2$ and $SD12$ can be mapped by simple formula.  The parameters have same functional dependence on lag number m suggesting similar source of variability of these quantities. The drastic reduction of the curvature of SD12 points out its significance in determining influence of regulatory system and environment and thus the health of the heart. The lagged Poincar\'{e} plot of RR-interval appears to be better suited for assessing heart rate variability.
Among the other analysis methods we have applied, we would like to emphasize on the novel application of autocorrelation technique to assess the autonomic regulation of a diabetic heart. To our knowledge this is the first attempt to differentiate between a normal and diabetic heart using autocorrelation technique. Results from the autocorrelation analysis of $RR_n$ interval suggest lack of short term variability of RR intervals among the diabetic patients. On the other hand, higher short term variability in normal subjects is marked by the presence of appropriate oscillation in correlation function. Diabetes as hypothesized impairs this variation and changes the RR interval distribution pattern. Moreover these change can be pictorially expressed in such a manner that even an untrained eye can distinguish the marked difference. These may prove to be really effective in an early diagnosis of diabetic autonomic neuropathy and may also have some prognostic importance.

%\begin{thebibliography}{letter}
\section{References}

\noindent [1] P. Contreras, R. Canetti2, E. R. Migliaro, Physiol Meas. (2007), 28(1), 85.

\noindent [2] U. R. Acharya, K. P. Joseph, N. Kannathal, C. M. Lim, J. S. Suri, Med Bio Eng Comput (2006), 44,1031-1051.

\noindent [3] H. Tsuji, M. G. Larson, F. J. Venditti, E. S. Manders, J. C. Evans, C. L. Feldman, D. Levy, Circulation(1996), 94, 2850-2855.

\noindent [4] J. Gerritsen, J.M. Dekker, B.J. TenVoorde , Diabetes Care (2001), 24, 1793-1798.

\noindent [5] Australian Diabetes Society (ADS) [http://www.diabetessoci ety.com.au/]

\noindent [6] S.C. Johnston, J.D. Easton, Stroke (2003), 4 ,2446.

\noindent [7] L. Duanping, M. Carnethon, W. E. Gregory, E. C Wayne, G. Heiss, Diabetes (2002), 51, 3524-3531.

\noindent [8] M. Kataoka, C. Ito, H. Sasaki, K. Yamaneb, N. Kohno, Diabetes Research and Clinical Practice (2004), 64 , 51-58.

\noindent [9] A. J. Kulinska, A. M. Grochowalska, K. Torzynska, L. Kramer, A. Sowinska, J. Moczko, T. Siminiak, Poznan Medical University, Poznan, Polan.

\noindent [10] D.J. Ewing, C.M. Martyn, R.J. Young, B.F. Clarke, Diabetes Care (1985), 8, 491-493.

\noindent [11] M. Pagani, Nutrition and Metabolism (2000), 13(6), 341-346.

\noindent [12] M.D. Rollins, J.G. Jenkins, D.J. Carson, B.G. McGlure, R.H. Mitchell,
S.Z. Imam, Diabetologia (1992), 35:452-455.

\noindent [13] A. H. Khandoker, H.F. Jelinek, M. Palaniswami, BioMedical Engineering OnLine (2009), 8, 3.

\noindent [14] V. Spallone, G. Menzinger, Diabetes (1997), 46 , S67.

\noindent [15] B. Goldstein and T. G. Buchman, J. Intensive Care Med (1998), 13, 252-65.

\noindent [16] M.P. Tulppo, T.H. Makikallio, T.E.S. Takala, T. Seppanen, Am J Physiol (1996), 271, H244-H252..

\noindent [17] M .Brennan, M .Palaniswami, P .Kamen, IEEE Trans on Biomed Engg (2001), 48, 1342-1347.

\noindent [18] M.A. Woo, W. G. Stevenson, D. K. Moser, R. B. Trelease and R. M Harper, Am. Heart J (1992), 123, 704-10.

\noindent [19] I. A. Khovanov, N. A. Khovanova, P. V. E. McClintock and A. Stefanovska,arXiv:0912.2237v1.

\noindent [20] M. Brennan, M .Palaniswami, P. Kamen, Am J Physiol Heart Circ Physiol (2002), 283, H1873-H1886.

\noindent [21] T. P .Thakre and M. L. Smith, BMC Cardiovasc Disorders (2006), 6, 27.

\noindent [22] C. Lerma, InfanteO, H.G. Perez and M.V.Jose, Clin. Physiol. Funct. Imaging (2003), 23, 72-80.

\noindent [23] A. Stefanovska and Bracic M Contemporary Phys. (1999), 40(1), 31-55.

\noindent [24] S.K.Ghatak, B.Roy, R.Choudhuri and Bandopadhaya, arXiv:1003.2075v1.

\noindent [25] C. K. Peng, S. Havlin, H .E. Stanley, A. L .Goldberger, Chaos, (1995), 5(1), 1054-1500.

\noindent [26] S. Julius and S. Nesbitt, Am J Hypertens (1996),  9, 113-120.

\noindent [27] F. Triposkiadis, G. Karayannis, G. Giamouzis, J. Skoularigis, G. Louridas, J. Butler, J Am Coll Cardiol ( 2009),  54, 1747-1762.

\newpage
\begin{figure}[htb]
\begin{center}
    \epsfig{file=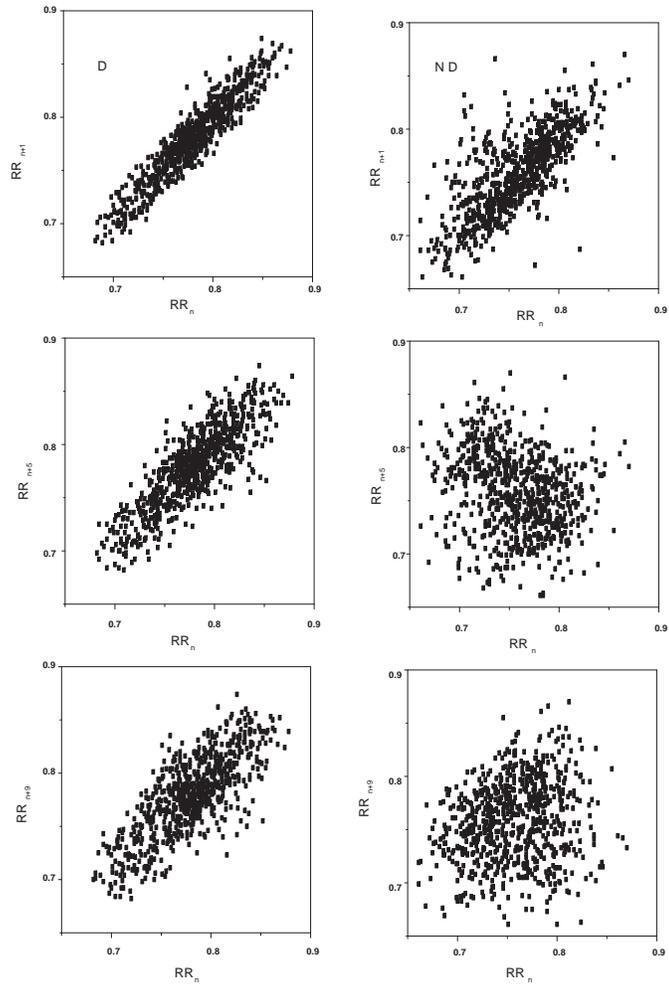,width=10cm}
    \caption{Poincare plot $RR_{n+m}$ vs $RR_n$ of one diabetic(D) and one non-diabetic (ND) subject of similar age for $M=1,5$ and $9$.}
\end{center}
\end{figure}
\newpage
\begin{figure}[htb]
\begin{center}
  \epsfig{file=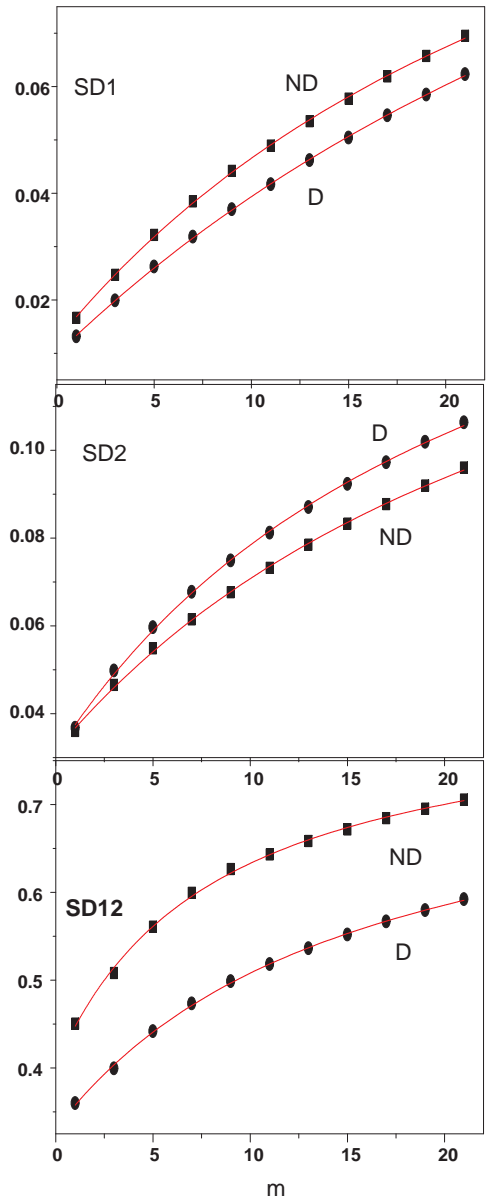,width=8cm}
   \caption{Variation of  mean $SD1$(upper),mean $SD2$ (middle) and mean $SD12$ (lower) with lag number $m$ for diabetic (D) and non-diabetic (ND) groups }
\end{center}
\end{figure}

\newpage
\begin{figure}[htb]
\begin{center}
    \epsfig{file=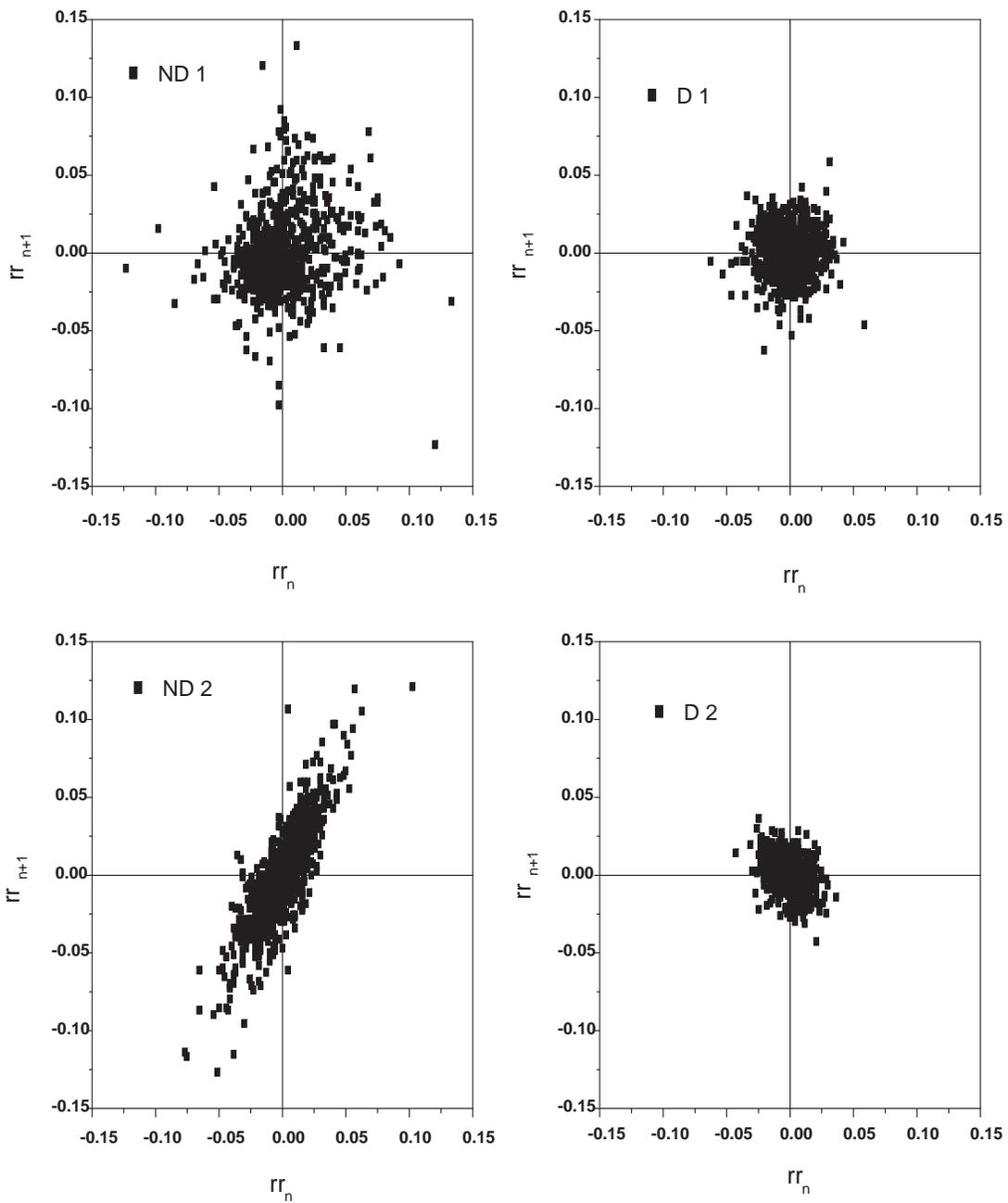,width=16cm}
    \caption{Plot $rr_{n+1}$ vs $rr_n$ of  diabetic(D) and  non-diabetic (ND) subjects of similar age.}
\end{center}
\end{figure}
\newpage
\begin{figure}[htb]
\begin{center}
    \epsfig{file=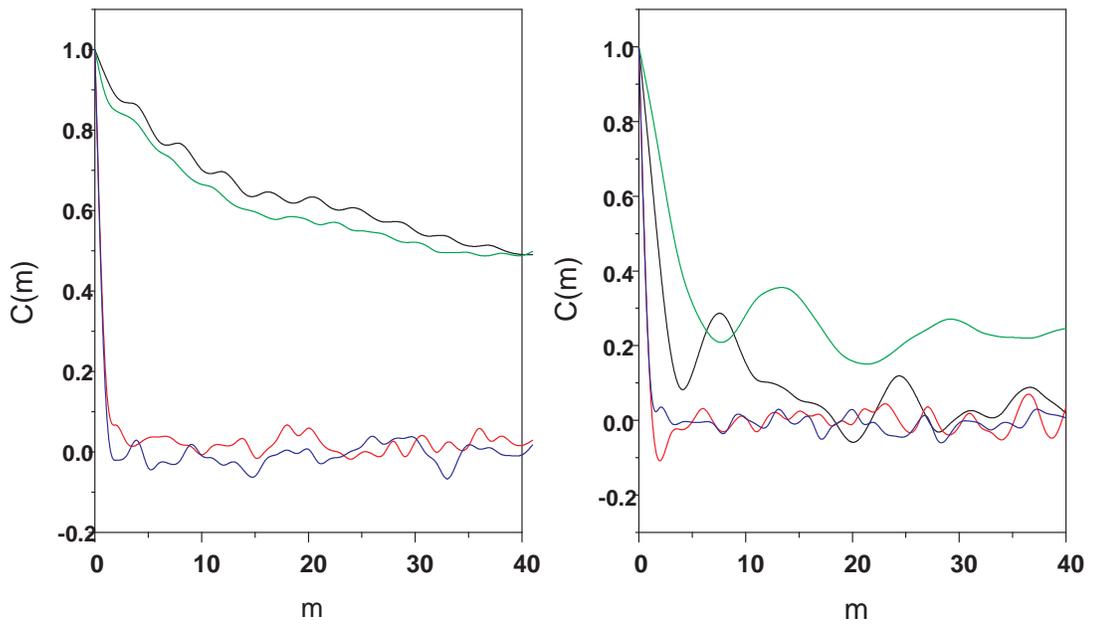,width=16cm}
   
    \caption{Plot $C(m)$ vs $m$  of  two diabetic(D1 and D2 ) and two non-diabetic (ND1 and ND2) subjects considered in Fig.3}
\end{center}
\end{figure}
\newpage
\begin{figure}[htb]
\begin{center}
    \epsfig{file=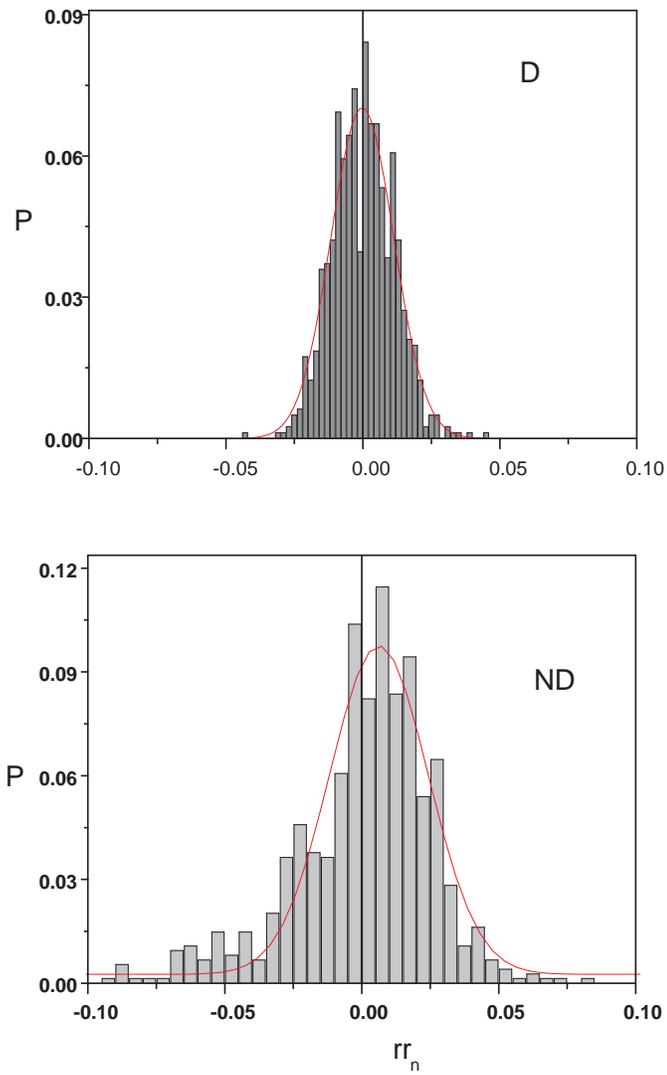,width=10cm}
    \caption{Plot of probability P as a function of $rr_n$ of  diabetic(D1) and  non-diabetic (ND1) subject of Fig.3}
\end{center}
\end{figure}
\newpage
\begin{figure}[htb]
\begin{center}
    \epsfig{file=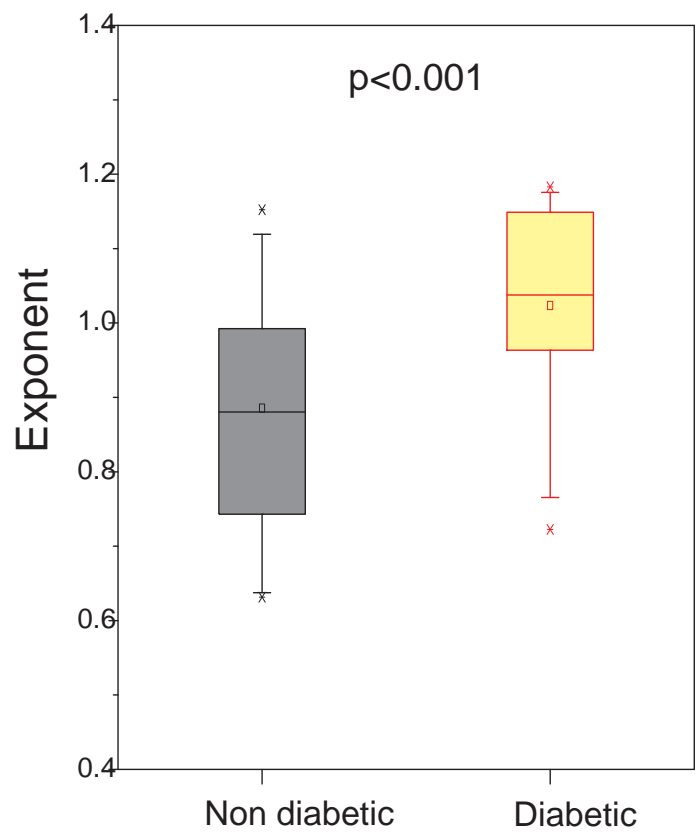,width=10cm}
    \caption{Exponent $\alpha$  for non-diabetic (ND) and diabetic(D) groups}
\end{center}
\end{figure}

\end{document}